\documentclass[11pt]{article}
\usepackage{latexsym,amsmath,amscd,amssymb,graphicx,amsbsy,color,xcolor,amsfonts,amsthm}

\begin{document}
\title{Synchronization in Uniformly Accelerated Frames}
\author{Tzvi Scarr\\
Jerusalem College of Technology\\
Department of Mathematics\\
P.O.B. 16031 Jerusalem 91160, Israel\\
e-mail: scarr@g.jct.ac.il}
%

\maketitle

\begin{abstract}
We show that a system is uniformly accelerated if and only if all of the clocks in the system can be synchronized to each other, and the clocks will remain synchronized as long as the acceleration remains uniform. In particular, it is possible to synchronize clocks on a disk rotating with constant angular velocity. Conventional thinking holds that this is impossible because a \emph{time gap} invariably arises.

\vskip0.2cm
 \textit{PACS}: 02.90.+p; 03.30.+p

 \textit{Keywords}:  uniform acceleration; clock synchronization; time dilation

\end{abstract}

\maketitle

\section{Introduction}\label{Intro}

We show here that a system is uniformly accelerated if and only if all of the clocks in the system can be synchronized to each other, and the clocks will remain synchronized as long as the acceleration remains uniform. In particular, it is possible to synchronize clocks on a disk rotating with constant angular velocity. Conventional thinking holds that this is impossible because a \emph{time gap} invariably arises.

The synchronization procedure presented here is based on the theory of covariant uniform acceleration \cite{FS1,FS2,FS3}. For the sake of completeness, we provide the necessary results from this theory in the next two sections. The synchronization procedure appears in section \ref{sync}.

\section{Covariant Uniform Acceleration}\label{bn}

In \cite{FS1}, \emph{uniform acceleration} was defined as a motion whose four-velocity $u(\tau)$ in an inertial system satisfies the equation
\begin{equation}\label{uam3}
c\frac{du^{\mu}}{d\tau}=A^{\mu}_{\;\;\nu}u^{\nu},
\end{equation}
where $\tau$ is the proper time of the accelerating object and $A_{\mu\nu}$ is a rank $2$ antisymmetric tensor whose components do not depend on $\tau$.
In the $1+3$ decomposition of Minkowski space, the acceleration tensor $A$ of equation (\ref{uam3}) has the form
\begin{equation}\label{aab2}
A^{\mu}_{\;\;\nu}=\left(\begin{array}{cc}0 & \mathbf{g}^T\\ &
\\\mathbf{g}&-c\mathbf{\Omega}^*\end{array}\right),
\end{equation}
where $\mathbf{g}$ is a 3D vector with physical dimension of acceleration, $\mathbf{\Omega}=(\Omega^1,\Omega^2,\Omega^3)$ is a 3D vector with physical dimension $1/\hbox{time}$, the superscript $T$ denotes matrix transposition, and
\[ \mathbf{\Omega}^*= c\varepsilon_{ijk}\Omega^k, \]
where $\varepsilon_{ijk}$ is the Levi-Civita tensor. The factor $c$ provides the necessary physical dimension of acceleration. The vector $\mathbf{g}$ represents \emph{linear} acceleration. If $\mathbf{\Omega}=0$, we obtain constant linear acceleration in a fixed direction, otherwise known as \emph{hyperbolic motion}. The vector $\mathbf{\Omega}$ is the angular velocity of the motion. If $\mathbf{g}=0$, we obtain \emph{pure rotational motion}.

In \cite{FS2}, we defined a \emph{frame} $K$ to be uniformly accelerated if there is a one-parameter family $\{K_\tau\}$ of inertial frames, instantaneously comoving to $K$, whose orthonormal bases $\{\lambda_{(0)}=u(\tau),\lambda_{(1)}(\tau),\lambda_{(2)}(\tau),\lambda_{(3)}(\tau)\}$ satisfy
\begin{equation}\label{uamlam2}
c\frac{d\lambda_{(\kappa)}^{\mu}}{d\tau}=A^{\mu}_{\;\;\nu}\lambda_{(\kappa)}^{\nu}\quad, \quad \kappa=0,1,2,3.
\end{equation}
This is known as generalized Fermi-Walker transport. Similar constructions can be found in \cite{mtw} and \cite{Hehl}. We stress that the solutions to (\ref{uamlam2}) satisfy Einstein's condition of \emph{constant acceleration in the comoving frame}.

\section{Velocity Transformations and Time Dilation}

The heart of the synchronization procedure lies in showing that the time dilation between any two given clocks at rest in $K$ is constant in time. In order to compute the time dilation, we first derive the velocity transformation from $K$ to the initial comoving frame $K_0$. For more details, see \cite{FS3}.

A particle's four-velocity in $K_0$ is, by definition, $\frac{dx^\mu}{d\tau_p}$, where $x(\tau_p)$ is the particle's worldline, and $\tau_p$ is the particle's proper time. However, from Special Relativity, it is known that the proper time of a particle depends on its velocity. In addition, it is known that the rate of a clock in an accelerated system also depends on its position, as occurs, for example, for linearly accelerated systems, as a result of gravitational time dilation. Thus, the quantity $d\tau_p$ depends on both the position and the velocity of the particle, that is, on the state of the particle.

Since we do not yet know the particle's proper time, it is not clear how to calculate the particle's four-velocity in $K_0$ directly from its velocity in $K$. To get around this problem, we will differentiate the particle's worldline by the proper time $\tau$ of the observer at the origin of $K$ instead of by $\tau_p$. We call the quantity
\begin{equation}\label{def4Dvel}
\tilde{u}=\frac{dx}{cd\tau}
\end{equation}
the \emph{4D velocity of the particle with respect to $\tau$}. The same technique was used by Horwitz and Piron \cite{offshell}, using the four-momentum instead of the four-velocity, thereby introducing the area known as ``off-shell" electrodynamics.
 
Since the 4D velocity is a tangent vector to the worldline, it is a scalar multiple of the four-velocity. Causality implies that this scalar is positive. Hence, the particle's four-velocity, which is a normalized tangent vector, is
\begin{equation}\label{4veland4dvel}
u=\frac{\tilde{u}}{|\tilde{u}|},
\end{equation}
the normalization of $\tilde{u}$.

We compute now a particle's four-velocity in $K_0$, given its 4D velocity in $K$. Let $y^{(\nu)}(\tau)$ be the worldline in the uniformly accelerated frame $K$ of a moving particle. Let $\tilde{w}^{(\nu)}=\frac{dy^{(\nu)}}{dy^{(0)}}$ denote the particle's 4D velocity in $K$ with respect to $\tau=c^{-1}y^{(0)}$.
As shown in \cite{FS3}, the particle's 4D velocity in $K_0$ at the point $\mathbf{y}$ of $K$, with respect to $\tau$, is
\begin{equation}\label{4velgen1}
 \tilde{u}=\frac{dx}{cd\tau}=\frac{dx}{dy^{(0)}} =(\tilde{w}^{(\nu)}+c^{-2}(A\bar{y})^{(\nu)})\lambda_{(\nu)}(\tau).
\end{equation}
On the other hand, denoting the proper time of the particle by $\tau_p$ and using the chain rule, we have
\begin{equation}\label{dtauyfromlenu}
u=\frac{dx}{d\tau_p}=\frac{dx}{d\tau}\frac{d\tau}{d\tau_p}=\tilde{u}\frac{d\tau}{d\tau_p}.
\end{equation}
Comparing (\ref{4veland4dvel}) and (\ref{dtauyfromlenu}), the time dilation between the particle and the observer at rest at the origin of $K$ is
\begin{equation}\label{dtauyequdtau}
\frac{d\tau}{d\tau_p}=\frac{1}{|\tilde{u}|}.
\end{equation}

To obtain the time dilation between a clock at rest at the point $\mathbf{y}$ of $K$ and the clock at rest at the origin of $K$, set
\begin{equation}\label{restclockw}
\tilde{w}=(1,0,0,0)
\end{equation}
in equation (\ref{4velgen1}). Then, since $\tilde{w},A$ and $\bar{y}$ are constant and the $\Lambda$'s are an orthonormal basis, the time dilation is constant.

\section{Synchronization}\label{sync}

Let $K$ be a uniformly accelerated frame, with acceleration tensor $A$. Our goal is to synchronize the clock $C$ at rest at the origin of $K$ with the clock $C_{\mathbf{y}}$ at rest at the arbitrary spatial point $\mathbf{y}$ of $K$. If this can be done for any point $\mathbf{y}$, then any two clocks at rest in $K$ can be synchronized to each other by composing two time dilations.

The proper time of $C$ is $\tau$, as is easily checked by substituting $\tilde{w}=(1,0,0,0)$ and $\bar{y}=(0,0,0)$ in (\ref{4velgen1}). Let $\tau_\mathbf{y}$ be the proper time of $C_{\mathbf{y}}$. From (\ref{4velgen1}) and (\ref{dtauyequdtau}), the time dilation $\frac{d\tau}{d\tau_{\mathbf{y}}}$ is
\begin{equation}\label{4velgen13}
 \frac{d\tau}{d\tau_\mathbf{y}}= \frac{1}{|(w^{(\nu)}+(c^{-2}A\bar{y})^{(\nu)})\lambda_{(\nu)}(\tau)|}.
\end{equation}
As mentioned in the previous section, for a given point $\mathbf{y}$, this time dilation is constant.

To synchronize $C$ with $C_{\mathbf{y}}$, we use the following procedure. Send a light signal back and forth from $C$ to $C_{\mathbf{y}}$ and set the time of $C_{\mathbf{y}}$ at the arrival of the signal to be equal to the average between the sent and received times of $C$. In other words, we use Einstein synchronization to synchronize the initial time of the clocks. Next, adjust the rate of $C_{\mathbf{y}}$ by multiplying it by the constant factor $\frac{d\tau}{d\tau_\mathbf{y}}$. The clocks $C$ and $C_{\mathbf{y}}$ will remain synchronized as long as $K$ maintains its uniform acceleration.

In particular, the above shows that it is possible to synchronize clocks in a rotating reference frame, as long as the rotational velocity is constant. In the literature (see \cite{Gron}, for example), it is claimed that this is impossible. The reason, however, that previous theories have failed to achieve synchronization on a rotating disk is that they employ the usual \emph{time dilation} of Special Relativity, which is appropriate for clocks moving with \emph{constant velocity}. In such a case, the time dilation is independent of the position of the clock. In \cite{FS3}, however, it is shown that in an \emph{accelerating} system, the time dilation depends on the position of the clock. Thus, the standard approach to clock synchronization breaks down for a system with acceleration, in particular, for a \emph{rotating} system.

\section{Discussion}\label{disc}

There are many interesting open questions about uniformly accelerated systems. Suppose, for example, that the frames $K'$ and $K''$ are both uniformly accelerated. Given an object's position, velocity, and acceleration in $K'$, what are its position, velocity, and acceleration in $K''$? Is $K''$ uniformly accelerated with respect to $K'$? Do the spacetime transformations between uniformly accelerated systems form a group?

\vskip0.7cm\noindent
The author would like to thank Y. Friedman for helpful suggestions and comments.


\end{document}